\documentclass[12pt,titlepage]{article}
\usepackage{amsmath}
\usepackage{amsfonts}
\usepackage{amssymb}
\usepackage{graphics}
\usepackage{epsfig}
\usepackage{color,amsmath,bm}
\usepackage{graphicx}

\newcommand {\slsh} [1] {\not{\hbox{\kern-2pt${#1}$}}}

\newcommand{\gsim}{\lower.7ex\hbox{$
\;\stackrel{\textstyle>}{\sim}\;$}}
\newcommand{\lsim}{\lower.7ex\hbox{$
\;\stackrel{\textstyle<}{\sim}\;$}}
\newcommand {\beq} {\begin{equation}}
\newcommand {\eeq} {\end{equation}}
\newcommand {\beqn}{\begin{eqnarray}}
\newcommand {\eeqn} {\end{eqnarray}}

\begin{document}

\begin{titlepage}

\begin{flushright}
FTPI-MINN-06/39, UMN-TH-2530/06\\
%February 19\\

\end{flushright}
\bigskip
\begin{center}
{\Large  {\bf The Hopf Skyrmion  in QCD with Adjoint\\[2mm] Quarks}
}
\end{center}
\bigskip
\begin{center}
{\large   S. {\sc Bolognesi}%\footnote{bolognesi@nbi.dk}
$^{(1)}$} and {\large M. {\sc Shifman}%\footnote{}
$^{(2)}$} \vskip
0.20cm
\end{center}
\begin{center}
$^{(1)}${\it \footnotesize The Niels Bohr Institute, Blegdamsvej 17, DK-2100
Copenhagen \O, Denmark}  \\ \vskip 0.20cm  $^{(2)}${\it \footnotesize
William I. Fine Theoretical Physics Institute, University of Minnesota, \\ 116
Church St. S.E., Minneapolis, MN 55455, USA}
\end {center}
\renewcommand{\thefootnote}{\arabic{footnote}}
\setcounter{footnote}{0}
\bigskip
\bigskip
\bigskip
\noindent
\begin{center}
{\bf Abstract}
\end{center}
We consider a modification of QCD in which conventional fundamental
quarks are replaced by Weyl fermions in the adjoint representation
of the color SU($N$). In the case of two flavors the low-energy
chiral Lagrangian is that of the Skyrme--Faddeev model. The latter
supports topologically  stable solitons with mass scaling as $N^2$.
Topological stability is due to the existence of a nontrivial Hopf
invariant in the Skyrme--Faddeev model. Our task is to identify, at
the level of the fundamental theory, adjoint QCD,
 an underlying reason  responsible for
the stability of the corresponding hadrons.
We argue that all ``normal" mesons and baryons, with mass
$O(N^0)$, are characterized by $(-1)^Q\, (-1)^F =1$,
where $Q$ is a conserved charge
corresponding to the unbroken U(1) surviving in the process
of the chiral symmetry breaking (SU(2)$\to$U(1) for two adjoint flavors).
Moreover, $F$ is the fermion number (defined mod 2 in the case at hand).
We argue that there exist exotic hadrons with mass
$O(N^2)$ and $(-1)^Q\, (-1)^F = -1$. They are in one-to-one correspondence with
the Hopf Skyrmions. The transition from nonexotic to exotic hadrons is due
to a shift in $F$, namely $F\to F-{\cal H}$
where ${\cal H}$ is the Hopf invariant. To detect this phenomenon we have to
extend the Skyrme--Faddeev model by introducing
fermions.

\end{titlepage}

\section{Introduction}

\label{intro}

In connection with the recent work on planar equivalence between
supersymmetric and non-supersymmetric gauge theories, and orientifold large-$
N$ expansion \cite{ASV}, there is a renewed interest in dynamics of
Yang--Mills theories with massless quarks in representations other than
fundamental. In particular, the orientifold large-$N$ expansion is based on
the fact that the quarks in the two-index antisymmetric representation at $
N=3$ are identical to conventional fundamental quarks.

If the number of flavors $N_f >1$, the theory under consideration has a
chiral symmetry which is spontaneously broken. The pattern of the chiral
symmetry breaking ($\chi$SB) depends on particular representation to which
the massless fermions of the theory belong \cite{mac,triangle,rm}. For $N_f$
Dirac fermions in the two-index antisymmetric representation the pattern of $
\chi$SB is identical to that of QCD, namely,
\begin{equation}
\mathrm{SU}(N_f)_L\times \mathrm{SU}(N_f)_R \to \mathrm{SU}(N_f)_V
\end{equation}

As a result, the low-energy limit is described by the same chiral Lagrangian
as in QCD, with the only distinction that in the case of two-index
antisymmetric quarks the pion constant $F_\pi$ scales as
\begin{equation*}
F_\pi^2 \sim N^2\,,
\end{equation*}
while in QCD it scales linearly (in QCD $F_\pi^2 \sim N$). As was noted in
\cite{Armoni:2003jk}, this seemingly minor difference leads to a crucial
consequence. Indeed, in both theories there exist Skyrmion solitons \cite
{Skyrme}. The issue of Skyrmions in QCD is well-studied. Witten showed \cite
{Witten,Witten:1983tx} that Skyrmions in QCD can be identified with baryons,
their topological charge being in one-to-one correspondence with the baryon
charge of the microscopic theory (for reviews see \cite{skyrmrev}). The
Skyrmion mass is proportional to $F_\pi^2$ and grows linearly with the
number of colors. This is fully compatible with our intuition since baryons
in multicolor QCD are composed of $N$ quarks. At the same time the very same
dependence of the Skyrmion mass on $F_\pi^2$ in the theory with the
two-index antisymmetric quarks implies that the Skyrmion mass grows as $N^2$
rather than $N$.

At first sight this is totally counterintuitive since in this case, just as
in QCD, one can construct a color-singlet baryon from $N$ antisymmetric
quarks.

This puzzle formulated in \cite{Armoni:2003jk} was successfully solved in
\cite{Bolognesi} (for follow-up works see \cite{Cohen}). It turns out that
if one considers $N$-quark colorless bound states, not all quarks can be in
the $S$-wave state, and, as a result, the mass of such objects scales as $
N^{7/6}$ \cite{Bolognesi}. The minimal number of quarks of which one
can build a particle with all quarks in the $S$-wave state is $N^2$.
This $N^2$ quark particle is the stable state described by the
Skyrmion. As for the $N$ -quark particles they are unstable with
respect to fusion of $N$ species into one $N^2$ quark state, with a
huge release of energy in the form of pion emission.

It is clear that with massless quarks in higher representations of the gauge
group, we deal with very interesting and intriguing dynamics calling for
further investigation. In this work we will continue in this direction. We
will consider $N_f $ massless \emph{Weyl} (or, which is the same, \emph{
Majorana}) fermions in the adjoint representation of the SU$(N)$ gauge
theory. These fermions will be referred to as ``quarks" or ``adjoint
quarks." To ensure a chiral symmetry in the fundamental Lagrangian on the
one hand, and to keep the microscopic theory asymptotically free on the
other, we must impose the following constraints on the value of $N_f$:
\begin{equation}
2\leq N_f\leq 5\,.  \label{fc}
\end{equation}
The pattern of the $\chi$SB in this case is also known \cite{mac,triangle,rm}
,
\begin{equation}
\mathrm{SU}(N_f)\times \mathbb{Z}_{2N N_f}\to \mathrm{SO}(N_f)
\times \mathbb{
Z} _{2} \,,  \label{pater}
\end{equation}
where the discrete factors are the remnants of the anomalous singlet axial
U(1). Below we will focus on the case $N_f=2$. A related discussion of the
three-flavor case is presented in Ref.~\cite{AuShi}.

Equation (\ref{pater}) can be elucidated as follows. If we denote the
adjoint quark field as $\lambda^a_{\alpha\, f}$ (here $a,\,\alpha ,\, f$ are
the color, Lorentz-spinorial, and flavor indices, respectively; we use the
Weyl representation), the Lorentz-scalar bilinear $\lambda^a_{\alpha\, f}\,
\lambda^{a\,\alpha}_{g}$ is expected to condense,
\begin{equation}
\langle\lambda^a_{\alpha\, f}\, \lambda^{a\,\alpha}_{g} \rangle \sim
\Lambda^3 \,.  \label{fivep}
\end{equation}
The above bilinear is obviously symmetric with respect to $f,g$; it is a
vector in the flavor space. It is convenient to choose the condensate in
such a way that the third component of the vector has a nonvanishing vacuum
expectation value,
\begin{equation}
\langle\lambda^a_{\alpha}\left( i\,\tau_3\,\tau_2\right)
\lambda^{a\,\alpha}\rangle = 2\,\Lambda^3\,,  \label{conde}
\end{equation}
where $\tau_i$ ($i=1,2,3$) stand for the Pauli matrices acting in the flavor
space (the same matrices acting in the spinor space will be denoted by $
\sigma_i$). Equation (\ref{conde}) implies that
\begin{equation}
\langle\lambda^a_{\alpha\, 1}\, \lambda^{a\,\alpha}_{2}\rangle =\Lambda^3\,,
\label{five}
\end{equation}
with other flavor components vanishing.

The above order parameter stays intact under those transformations from SU$
(2)$ which generate rotations around the third axis in the flavor space.
There is one such transformation,
\begin{equation}
\lambda \to e^{i\, \alpha\,\tau_3}\, \lambda\,.  \label{six}
\end{equation}
Thus, the low-energy pion Lagrangian is a nonlinear sigma model with the
target space $\mathcal{M}$ given by the coset space
\begin{equation}
\mathcal{M}_2 = \mathrm{SU}(2)/ \mathrm{U}(1) =S^2\,.
\end{equation}
This is the famous O(3) sigma model which was treated in two
dimensions in \cite{Polyakov}. In four dimensions, with quartic in
derivatives terms included (for stabilization), this is the model
\cite{Faddeev} which is sometimes referred to as the Skyrme--Faddeev
and sometimes as the Faddeev--Hopf model. Soliton's topological
stability is due to the existence of the Hopf invariant. In fact,
the Faddeev--Niemi solitons are knots \cite {Faddeev}. For generic
values of $N_f$ the sigma model is defined on the coset
space\,\footnote{ This is also true for SO($N)$ Yang--Mills theories
with vectorial quarks in which the structure of the chiral
Lagrangians and solitons is similar. In this case the Skyrmion was
identified with the baryon \cite{Witten:1983tx}. Its
 $\mathbb{Z}_2$ stability is due to the fact that the $\epsilon$
tensor cannot be written as the sum of $\delta$'s.}
\begin{equation}
\mathcal{M}_{N_f} = \mathrm{SU}(N_{f})/\mathrm{SO}(N_{f})\,.  \label{targ}
\end{equation}
The starting point of the present work is the observation that the third
homotopy group for (\ref{targ}) is nontrivial in all four cases (\ref{fc}),
as shown in Table~\ref{tabone}.

\begin{table}[tbp]
\begin{center}
\begin{tabular}{|c|c|c|c|c|}
\hline
\rule{0mm}{5mm} & $N_{f}=2$ & $N_{f}=3$ & $N_{f}=4$ & $N_{f}=5$ \\
[1mm] \hline
\rule{0mm}{5mm} $\pi_{3}$ & $\mathbb{Z}$ & $\mathbb{Z}_{4}$ &
$\mathbb{Z} _{2} $ & $\mathbb{Z} _{2}$ \\[1mm] \hline
\end{tabular}
\end{center}
\caption{{\protect\footnotesize The third homotopy group for sigma models
emerging in Yang--Mills with two, three, four and five adjoint flavors.}}
\label{tabone}
\end{table}

Hence, topologically stable solitons (whose mass scales as $N^2$) exist much
in the same way as Skyrmions in QCD. Unlike QCD, where the relation between
the Skyrmions and microscopic theory is well established, in our case it is
far from being clear. How can one interpret the topologically stable
solitons of the low-energy effective Lagrangian from the standpoint of the
fundamental theory with the adjoint quarks? With regards to stability, is it
an artifact of the low-energy approximation? If no, what prevents these
particle whose mass scales as $N^2$ from decaying into ``light"
color-singlet mesons and baryons with mass $O(N^0)$? What plays the role of
the QCD baryon charge?

Below we will answer these and similar questions. We will argue that the
Hopf Skyrmion stability is not a low-energy artifact; rather it can be
understood in the fundamental theory -- Yang--Mills with two adjoint quarks.
This is due to the fact that all conventional mesons and baryons with $
m=O(N^0)$ in the theory at hand have
\begin{equation}
(-1)^Q\cdot (-1)^F =1\,,
\end{equation}
while for the Hopf Skyrmion
\begin{equation}
(-1)^Q\cdot (-1)^F =-1\,.
\end{equation}
Here $Q$ is the conserved global U(1) charge which survives after the
spontaneous breaking (\ref{five}), and $F$ is the fermion number. Note that
the fermion number \emph{per se} is not conserved in the microscopic theory
under consideration, but $(-1)^F$ is.

Organization of the paper is as follows. Section \ref{general} summarizes
known facts about QCD with adjoint quarks as well as our findings regarding
the existence of exotic stable hadrons with mass $O(N^2)$ in the two-flavor
model. In Sect. \ref{skyrme} we discuss the low-energy limit of two-flavor
adjoint QCD. We present the corresponding chiral Lagrangian and discuss its
features. In Sect.~\ref{massiv} we extend the model by introducing
appropriate fermion fields. In Sect.~\ref{impa} we calculate the induced
fermion charge.

\section{General considerations}

\label{general}

The SU(2) flavor group is broken down to the U(1) subgroup generated by the
Pauli matrix $\tau_{3}$. The unbroken U(1) is the exact global symmetry of
the microscopic theory acting on the adjoint quark field as indicated in
Eq.~(\ref{six}). Correspondingly, the U(1) charge (to be referred to as $Q$)
of the quark of the first flavor is $+1$ while that of the second flavor is $
-1$. It is obvious that the condensate (\ref{five}) is neutral.

There are two Nambu--Goldstone bosons, $\pi^{++}$ and $\pi^{--}$. Roughly, $
\pi^{++}\sim \lambda_1\lambda_1\,,\,\, \bar\lambda^2\bar\lambda^2$ and $
\pi^{--}\sim \lambda_2\lambda_2 \,,\,\, \bar\lambda^1\bar\lambda^1$. (Note
that complex conjugation raises the flavor index). Two of four possible
linear combinations produce massive scalar mesons $\sigma$ while two others
produce massless pseudoscalar pions. Note that, although we use the Weyl
formalism for the fermions, parity is conserved in the microscopic theory
under consideration at any $N_f$.

The pion U(1) charges are
\begin{equation}
Q(\pi^{\pm\pm} ) = \pm 2\,.
\end{equation}
If we rely on a constituent quark model, as we do in QCD, we could say that
for a generic hadronic state
\begin{equation}
Q = 2\,J\,\,\, (\mathrm{mod} \,\, 2)\,,  \label{QJ}
\end{equation}
where $J$ is the spin of the given hadron. Alternatively this relation can
be represented as
\begin{equation}
(-1)^Q = (-1)^{2J}\,.
\end{equation}
\mbox{} \vspace{2mm} \mbox{}

The current corresponding to the conserved U(1) charge in our conventions
can be written as
\begin{eqnarray}
&& J^{\mathrm{U}(1)}_\mu \equiv \frac{1}{2} \left(\sigma_\mu\right)_{\alpha
\dot\alpha} J^{\mathrm{U}(1),\,\, \alpha\dot\alpha}\,,  \notag \\[3mm]
&& J^{\mathrm{U}(1),\,\, \alpha\dot\alpha}= \bar{\lambda}^{a\,\,\dot\alpha\,
\,\, 1}\,{\lambda}^{a\,\,\alpha}_1 - \bar{\lambda}^{a\,\,\dot\alpha\,\,\,
2}\,{\lambda}^{a\,\,\alpha}_2\,.
\end{eqnarray}
The second linear combination,
\begin{equation}
J^{F,\,\, \alpha\dot\alpha}= \bar{\lambda}^{a\,\,\dot\alpha\,\,\, 1}\, {
\lambda}^{a\,\,\alpha}_1 + \bar{\lambda}^{a\,\,\dot\alpha\,\,\, 2}\, {\lambda
}^{a\,\,\alpha}_2  \label{tue-one}
\end{equation}
can be called the fermion current. Classically it is conserved too. However,
at the quantum level, due to the chiral anomaly,
\begin{equation}
\partial^\mu\, J^{F}_\mu = \frac{NN_f}{16\pi^2}\, F_{\mu\nu}^a\, \tilde{F}
^{a\,\mu\nu}\,.  \label{aneq}
\end{equation}
For two adjoint flavors $N_f=2$ in Eq.~(\ref{aneq}). Thus, the fermion
number $F$ is not conserved. A discrete subgroup $\mathbb{Z}_{2N N_f}$
survives from (\ref{aneq}). The nonvanishing condensate (\ref{conde}) breaks
$\mathbb{Z}_{2N N_f}$ down to $\mathbb{Z}_{2}$. This means that
nonconservation of $F$ is quantized, $|\Delta F| =2$. In other words, $(-1)^F
$ is conserved.

For all ``ordinary" hadrons which can be produced from the vacuum by local
currents, say, (\ref{tue-one}) or (\ref{mafe}), determination of $(-1)^F$ is
straightforward. This is not the case, however, for hadrons build of $\sim
N^2$ constituents, see below.

Now, we can decompose the Hilbert space of hadronic excitations in the
direct sum of two spaces
\begin{equation}
\mathcal{H}^{\left(\mathrm{hadronic}\right)}=\mathcal{H}^{\left(+1,+1\right)
}\oplus\mathcal{H} ^{\left(-1,-1\right) } \label{Hilbert senza
fractional}
\end{equation}
containing, respectively, the composite states with the even and odd
U(1) charges. We have denoted the charges as
$\left((-1)^{Q},(-1)^{F}\right)$. From the point of view of the
hadronic Hilbert space (\ref{Hilbert senza fractional}) this would
appear a redundant notation. It will soon be clear that this is not.

Relying on the same constituent quark model for orientation, we
would say that $\mathcal{H}^{\left( +1,+1\right)} $ contains
hadronic excitation of the boson type while $\mathcal{H}
^{\left(-1,-1\right)}$ of the fermion type.  In particular,
$\mathcal{H} ^{\left( +1,+1\right) }$ contains the massless
Nambu--Goldstone bosons $\pi^{\pm\pm}$ and, hence, there is no mass
gap here. On the contrary, $\mathcal{H}^{\left(-1,-1\right) }$ has a
mass gap $m$, the mass of the lightest composite fermion of the type
\begin{equation}
\psi_{\beta\,\,f}= C\, \mathrm{Tr}\left( \lambda^{\alpha}_f\,F_{\alpha{\beta}
}\right) \equiv C\, \mathrm{Tr}\left( \lambda^{\alpha}_f \sigma_{\alpha{\beta
} }^{\mu\nu}F_{\mu\nu}\right),  \label{mafe}
\end{equation}
were $F_{\alpha\beta}$ is the (anti)self-dual gluon field strength tensor
(in the spinorial notation), and $C$ is a normalizing factor,
\begin{equation*}
C\sim (N\Lambda^2)^{-1}\,.
\end{equation*}
Two U(1)-charge $\pm 1$ composite fermions are
\begin{equation}
\psi_\pm =C\,\mathrm{Tr}\left( \lambda_{1,2}^{\alpha}\,
F_{\alpha{\beta} }\right) \,, \label{fundamentalfermion}
\end{equation}
(plus their antiparticles, of course). Note that $\psi_-$ is \emph{not} $
\psi_+$'s antiparticle. Moreover, we can combine $\psi_{+}$ and $
\bar\psi_{-} $ in a single Dirac spinor $\Psi_D$,
\begin{equation}
\Psi_D =\left\{
\begin{array}{c}
\psi_{1} \\[1mm]
- i\, \sigma_2 \, \left({\psi}_{2}\right)^*
\end{array}
\right\} .  \label{fri-three}
\end{equation}
Here $\sigma_2$ lowers the spinorial dotted index. This construction is
useful for comparison of our results with calculations one can find in the
literature, see e.g. Ref.~\cite{lfan}. Although this work is intended for
description of somewhat different (albeit related) physics, its mathematical
aspect overlaps with that of our analysis.\footnote{
For a supersymmetric follow up see \cite{lf}.}

Below we will argue that in fact Eq.~(\ref{Hilbert senza fractional}) is
incomplete. An extra sector can and must be added,
\begin{equation}
\mathcal{H}=\mathcal{H}^{\left(\mathrm{hadronic}\right)}
\oplus\mathcal{H}^{\left( \mathrm{exotic} \right) } \,,
\label{extras}
\end{equation}
where $\mathcal{H}^{\left(\mathrm{hadronic}\right)}$ is given by
(\ref{Hilbert senza fractional}) and the new sector is given by
\begin{equation}
\mathcal{H}^{\left(\mathrm{exotic}\right)}=\mathcal{H}^{\left(+1,-1\right)
}\oplus\mathcal{H} ^{\left(-1,+1\right) }\ .
\end{equation}
$\mathcal{H}^{\left( \mathrm{exotic}\right) }$ includes hadrons with
even $Q$ and odd $F$ and vice versa, odd $Q$ and even $F$. To build
such a hadron one needs $\sim N^2$ constituents. In adjoint QCD they
play the role of baryons of conventional QCD. Their existence is
reflected in the Hopf-Skyrmions.

\section{The Skyrme--Faddeev model}

\label{skyrme}

Now, let us briefly review the effective low-energy pion Lagrangian
corresponding to the given pattern of the $\chi$SB, see Eq. (\ref{pater})
with $N_f=2$. We describe the pion dynamics by the $O(3)$ nonlinear sigma
model (in four dimensions)
\begin{equation}
\mathcal{L}_{\mathrm{eff}}=\frac{F_{\pi}^{2}}{2}\,\, \partial_{\mu}\vec{n}
\cdot\partial^{\mu}\vec{n} +\mbox{higher derivatives}
\label{senza il fermione}
\end{equation}
where the three-component field $\vec n$ is a vector in the \emph{flavor}
space subject to the condition
\begin{equation}
\vec n^{\,2} =1\,,  \label{ts}
\end{equation}
and a higher derivative term is needed for stabilization. The ``plain"
vacuum corresponds to a constant value of $\vec n$ which we are free to
choose as $\langle n_3\rangle =1.$ Due to (\ref{ts}) the target space of the
sigma model at hand is $S^2$.

Usually the higher derivative term is chosen as follows (for a review see
\cite{manton}):
\begin{equation}
\delta \mathcal{L}_{\mathrm{eff}} = -\frac{\lambda}{4}\,
\left(\partial_{\mu} \vec{n}\times\partial_{\nu}\vec{n} \right)
\cdot\left(\partial^{\mu}\vec{n} \times\partial^{\nu}\vec{n} \right)\,.
\label{senza il fermionep}
\end{equation}
Equations (\ref{senza il fermione}) and (\ref{senza il fermionep})
constitute the Skyrme--Faddeev (or the Faddeev--Hopf) model. Note that the
Wess--Zumino--Novikov--Witten (WZNW) term \cite{wznw} does not exist in this
model.

To have a finite soliton energy the vector $\vec n$ for the soliton solution
must tend to its vacuum value at the spatial infinity,
\begin{equation}
\vec n\to \{0,0,1\}\,\,\,\mbox{at}\,\,\, \left| \vec x \right| \to\infty\,.
\label{vac}
\end{equation}
Two elementary excitations near the vacuum $n_3 =1$,
\begin{equation*}
\frac{1}{\sqrt 2}\left(n_1\pm i\, n_2\right),
\end{equation*}
can be identified with the pions. The boundary condition (\ref{vac})
compactifies the space to $S^3$. Since $\pi_3 (S^2) = \mathbb{Z}$, see Table~
\ref{tabone}, solitons present topologically nontrivial maps of $S^3\to S^2$
. As was noted in \cite{Faddeev}, there is an associated integer topological
charge $N$, the Hopf invariant, which presents the soliton number. This
charge cannot be the degree of the mapping $S^3\to S^2$ because dimensions
of $S^3$ and $S^2$ are different.

\begin{figure}[th]
\begin{center}
\leavevmode \epsfxsize 8.5 cm \epsffile{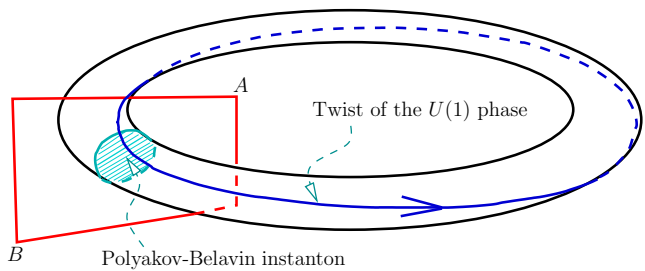}
\end{center}
\caption{{\protect\footnotesize The simplest Hopf soliton, in the adiabatic
limit, correspond to a Belavin-Polyakov soliton closed into a donut after a $
2\protect\pi$ twist of the internal phase.}}
\label{donu}
\end{figure}
\begin{figure}[th]
\begin{center}
\leavevmode \epsfxsize 11 cm \epsffile{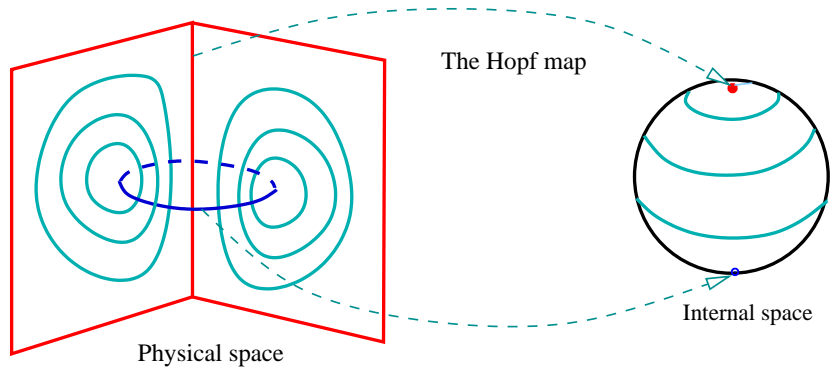}
\end{center}
\caption{{\protect\footnotesize The Hopf map from the space $R^3$ to
the sphere $S^2$.}} \label{hopfmap}
\end{figure}

The solitons in the the Skyrme--Faddeev model are of the knot type. The
simplest of them is toroidal, it looks as a ``donut," see e.g.~\cite{manton}
. Qualitatively it is rather easy to understand, in the limit when the ratio
of the periods is a large number, that the Hopf topological number combines
the instanton number in two dimensions, with twist in the perpendicular
third dimension.

Let us slice the ``donut" soliton by a perpendicular plane $AB$. In
the vicinity of this plane the soliton can be viewed as a cylinder,
so that the problem becomes effectively two-dimensional. In two
dimensions the O(3) sigma model has the Polyakov--Belavin instantons
\cite{Polyakov} whose topological stability is ensured by the
existence of the corresponding topological charge. The
Polyakov--Belavin instanton has an orientational collective
coordinate describing its rotation in the unbroken U(1) subgroup
(for a review see \cite{nsvz}). In two dimensions for each given
instanton this collective coordinate is a fixed number. In the Hopf
soliton of the type shown in Fig.~\ref{donu}, as we move the plane
$AB$ in the direction indicated by the arrow, this collective
coordinate changes (adiabatically), so that the $2\pi$ rotation of
the plane $AB$ in the direction of the arrow corresponds to the
$2\pi$ rotation of the orientational modulus of the
Polyakov--Belavin instanton. This is the twist necessary to make the
Hopf soliton topologically stable.

Another way to understand the geometry of the Hopf map is given in
Fig.~\ref{hopfmap}. Let us visualize the space $R^3$  as a ``book.''
Every ``page'' of this book is a semi-infinite plane attached to the
axial line. The axial line plus the points at infinity are mapped onto
the noth pole of the target space $S^2$. The preimage of the
south pole is a circle linked with the axial line. Every semi-infinite
plane is wrapped around the target space. The  U(1)  phase, which in
this picture is the rotation of $S^2$ that keeps fixed the north and
 south poles, is twisted while the semi-infinite plane is rotated
around the axial line.

The Hopf charge cannot be written as an integral of any density
which is local in the field $\vec n$. However, if one uses a U(1)
gauged formulation of the CP(1) sigma model in terms of the doublet
fields $z^\alpha,\,\, \bar z_\alpha $ ($\alpha = 1,2$) and
\begin{equation}
\bar z_\alpha\, z^\alpha =1\,, \qquad \vec n = -\bar z \, \vec \tau\, z\,,
\label{home}
\end{equation}
(for a review see \cite{ewi}), in this formulation the Hopf invariant
reduces to the Chern--Simons term for the above gauge field \cite{Faddeev},
\begin{eqnarray}
\mathcal{H} &=& \frac{1}{4\pi ^{2}}\int d^{3}x\,\epsilon ^{\mu \nu \rho }
\left( A_{\mu}\partial _{\nu }A_{\rho }\right), \\[3mm]
A_\mu &=& -\frac{i}{2}\, \bar z \,\overset{\leftrightarrow}{\partial}_\mu\, z
\label{tue-two}
\end{eqnarray}
(see Sect.~\ref{massiv}). Spherically symmetric field configurations
automatically have vanishing Hopf charge.

One can ask whether the Hopf Skyrmion represents a hadron with integer or
semi-integer spin. Quantization of the Hopf solitons was considered in the
literature previously \cite{Glad,Krus}, with the conclusion that it can be
quantized both as a boson and as a fermion. The possibility of the fermionic
quantization is due to the Finkelstein--Rubinstein mechanism \cite{FR}. In
both cases Eq.~(\ref{QJ}) is valid.

Since we would like to understand the relation between $Q$ and $F$ we need
to introduce (composite) fermions in the Skyrme--Faddeev model. The
low-energy description in terms of the $\vec n$ field knows nothing about
baryons of the type (\ref{mafe}). It is clear that it is impossible to
decide this issue without including such fermions in the low-energy chiral
model. This is done in Sect.~\ref{massiv}.

\section{Introducing massive fermions}

\label{massiv}

The Hopf soliton is topologically stable due to nontriviality of the
homotopy group $\pi_{3}\left( S^{2}\right) =\mathbb{Z}$. However, the
Skyrme--Faddeev model neglects excitations belonging to the odd sector of
the Hilbert space, baryons of the type (\ref{mafe}), whose mass is $
O(\Lambda)$ and does not scale with $N$. Let us now switch them on. One can
take baryons into account through the SU(2) invariant coupling $\psi^{\alpha
f} \vec n\cdot \left(\vec\tau\right)^g_f\,\psi_{\alpha g}$ + h.c. We can
write an effective Lagrangian that includes both the pions and the baryons,
\begin{equation}
\mathcal{L}_{\mathrm{eff}}= \frac{F_{\pi }^{2}}{2} \, \partial _{\mu } \vec{n
}\cdot \partial ^{\mu }\vec{n}+ \bar\psi^{ f}_{\dot\alpha} \,
i\,\partial^{\dot\alpha\alpha} \,\psi_{\alpha f} -\frac{g}{2}\left\{
\psi^{\alpha f} \vec n\cdot \left(\vec\tau\right)^g_f\,\psi_{\alpha g} +
\mathrm{h.c.}\right\} +\dots
\label{lagrangiana}
\end{equation}
where $g$ is a coupling constant of dimension of mass (which will be assumed
to be positive). Similar (but not identical) couplings were considered in
the literature previously, see e.g. Ref.~\cite{Jaroszewicz:1984xw}. The
Lagrangian (\ref{lagrangiana}) is obviously SU(2) invariant. The $\psi$
baryon mass is generated in the process of $\chi$SB. Namely, if in the
vacuum $n_3 =1$, the baryon term becomes
\begin{equation}
\mathcal{L}_{\mathrm{ferm}} =\bar\psi^{ f}_{\dot\alpha} \,
i\,\partial^{\dot\alpha\alpha} \,\psi_{\alpha f} -{g}\left\{ \psi^{\alpha
}_1 \,\psi_{\alpha \,2} +\mathrm{h.c.}\right\} \,,
\label{lagrangianslowlyp}
\end{equation}
(cf. Eq.~(\ref{conde})) and the mass $M^2_\psi=g^2$. The bilinear pion terms
are
\begin{equation}
\mathcal{L} = \left( \partial _{\mu } \pi ^{++}\right)\left( \partial^{\mu
}\pi ^{--}\right)\, .  \label{lagrangianslowly}
\end{equation}
Expanding near $n_3 =1$ we get interactions terms. For what follows it is
convenient to transfer the $\psi \psi\, n$ interaction from the potential to
the kinetic term. To this end let us introduce an SU(2) flavor matrix $U$
such that
\begin{equation}
U^\dagger\, (\vec n\cdot \vec\tau )\, U =\tau_3\,,
\end{equation}
where $U$ is a function of $\vec n$. It is not uniquely defined, of course.
One can always choose it in the form
\begin{equation}
U = \exp\left( i \vec \nu\cdot \vec\tau_\perp \right)\,,\qquad \nu_i \propto
\varepsilon_{ij}\, n_{\perp j}\,.
\end{equation}
Here $\vec\nu$ and $\vec n_\perp$ are two-dimensional vectors,
\begin{equation*}
\vec\nu =\{\nu_1,\,\nu_2\}, \quad \vec n_\perp =\{n_1,\,n_2\}
\end{equation*}
while $\vec\tau_\perp = \{\tau_1,\,\tau_2\}$. Furthermore, if
\begin{equation}
\psi_f \equiv U_f^g\, \chi_g\,,\qquad \psi^f \equiv \chi^g
\left(U^\dagger\right)^f_g\,
\end{equation}
the fermion part of the Lagrangian takes the form
\begin{eqnarray}
\mathcal{L}_{\mathrm{ferm}} &=& =\bar\chi_{\dot\alpha} \left(
i\,\partial^{\dot\alpha\alpha} + \mathcal{A}^{\dot\alpha\alpha} \right)
\chi_\alpha -{g}\left\{ \chi^{\alpha }_1 \,\chi_{\alpha \,2} +\mathrm{h.c.}
\right\} \,,  \label{fri-two} \\[4mm]
\mathcal{A} _{\mu } &\equiv & i\,U^\dagger \partial_\mu\, U \equiv \mathcal{A
} _{\mu }^{(i)}\,\tau_i\,.  \label{cala}
\end{eqnarray}
Written in terms of $\chi$'s the interaction terms have a generic form $
\bar\chi \, n \,...\, \partial n \, ...\, n \,\chi$.

The fermion current we are interested in is
\begin{equation}
J^{F0}_{\alpha\dot\alpha}= \bar{\chi}^{f}_{\dot\alpha}\,\,{\chi}_{\alpha
\,\, f}\,,\qquad F=\int\,d^3 x\, J^0  \label{fri-one}
\end{equation}
(cf. (\ref{tue-one})). The meaning of the superscript 0 will become clear
shortly. The current (\ref{fri-one}) is flavor-singlet since the summation
over $f$ is implied. Naively, one would say that states containing odd
number of $\chi$'s have $F=1,$ mod 2, while those with even number of $\chi$
's (in particular, no $\chi$'s at all) have $F=0,$ mod 2. The mass term in
Eq.~(\ref{fri-two}) satisfies the constraint $|\Delta F|=2$.

The question we ask is whether a state build from $n$'s can have $F=1,$ mod
2, and vice versa, a state containing a $\chi$ quantum $F=0,$ mod 2. Such
states cannot be produced from the vacuum by local sources constructed from
(a finite number of) $\chi$ and $n$ fields. We want to show, however, that
they do exist in the sector with nontrivial values of the Hopf invariant.

To see that this is the case let us examine the divergence of $J^{F0}_\mu$.
Naively one would say that $\partial^\mu J^{F0}_\mu$ is given by the
classical mass term in Eq. (\ref{fri-two}). A closer inspection reveals an
anomaly. Indeed, the U(1) symmetry corresponding to rotations around the
third axis in the flavor space is a strictly conserved global symmetry in
the model at hand. Maintaining conservation of the corresponding current
produces an anomalous contribution in $\partial^\mu J^{F0}_\mu$, as it
usually happens with the triangle graphs. Calculation of the anomaly is a
straightforward task, given that we keep only $\mathcal{A} _{\mu }^{(3)}$ in
Eq.~(\ref{fri-two}) treating $\mathcal{A} _{\mu }^{(3)}$ as an external
field,
\begin{equation}
\partial^\mu J^{F0}_\mu =\frac{1}{8\pi^2}\, F_{\mu\nu}^{(3)}\, {\tilde F}
_{\mu\nu}^{(3)}+\mbox{classical term}\,,
\label{anom40}
\end{equation}
where
\begin{equation}
F_{\mu\nu}^{(3)} = \partial_\mu \mathcal{A}_\nu^{(3)} -\partial_\nu \mathcal{
A}_\mu^{(3)}\,,
\end{equation}
and the extra $1/2$ in the coefficient ($8\pi^2$ in the denominator instead
of $4\pi^2$) reflects the fact that $\chi$ is the Weyl rather than Dirac
spinor.

Now let us take into account the fact that
\begin{equation}
F_{\mu\nu}^{(3)}\, {\tilde F}_{\mu\nu}^{(3)} =2\,\partial_\mu\,
K^\mu\,,\qquad K^\mu =\epsilon^{\mu\nu\alpha\beta}\,
A_\nu^{(3)}\,\partial_\alpha\, A_\beta^{(3)}\,.
\end{equation}
Here $K^\mu$ is the Chern-Simons current. Thus, we see that the genuine
fermion current is
\begin{equation}
J^{F}_\mu = J^{F0}_\mu -\frac{1}{4\pi^2}\, K_\mu  \label{fri-four}
\end{equation}
rather than $J^{F0}_\mu $. The fermion charge $F$ is shifted accordingly.

\section{The impact of $K_\protect\mu$}

\label{impa}

Consider first the Skyrme--Faddeev theory in $2+1$ dimensions. The conserved
topological current of the Belavin--Polyakov (BP) instanton \cite{Polyakov}
is
\begin{equation}
j_{\mathrm{BP}}^{\mu }=\frac{1}{8\pi }\, \epsilon ^{\mu \nu \rho }\epsilon
^{abc}n^{a}\partial _{\nu }n^{b}\partial _{\rho }n^{c}~,  \label{bpc}
\end{equation}
normalized so that $\int d^{2}xj_{\mathrm{BP}}^{0}=1$ on the instanton of
topological charge $1$ (see e.g. \cite{nsvz}, Sect. 6).

\begin{figure}[th]
\begin{center}
\leavevmode \epsfxsize 6.2 cm \epsffile{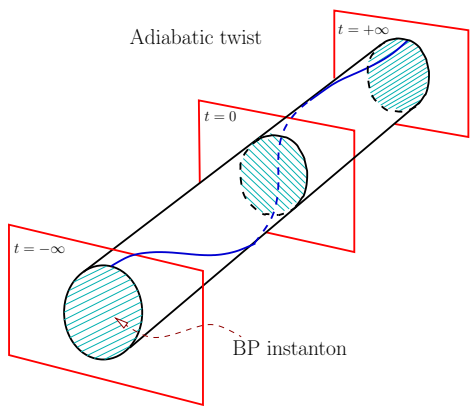}
\end{center}
\caption{{\protect\footnotesize A 2+1 dimensional (cylindrical) field
configuration with an adiabatically twisted BP instanton corresponding to
the unit Hopf invariant.}}
\label{bof}
\end{figure}

The conservation of $j_{\mathrm{BP}}^{\mu }$ is due to the condition $\vec
n^{\,2} =1$. Since $\partial _{\mu }j_{\mathrm{BP}}^{\mu }=0$, we can
introduce an auxiliary field $a_{\mu }$ defined as follows:
\begin{equation}
j_{\mathrm{BP}}^{\mu }=\epsilon ^{\mu \nu \rho }\, \partial _{\nu }a_{\rho
}\,.  \label{aintroduced}
\end{equation}
Then, the Hopf invariant can be expressed as \cite{Wilczek:1983cy}
\begin{equation}
\mathcal{H=}\int d^{3}x\, \epsilon ^{\mu \nu \rho }\, a_{\mu }\partial
_{\nu}\, a_{\rho }\,.  \label{hopf}
\end{equation}
It correspond to a Chern-Simon term\,\footnote{
If a term $\theta \mathcal{H}$ is added to the action, the BP particle
acquires a spin equal to ${\theta }/{2\pi }$ \cite{Wilczek:1983cy}.} for the
auxiliary gauge field $a_{\mu }$. The Hopf invariant is normalized in such a
way that it is $1$ on the BP field configuration of topological charge $1$
which adiabatically rotates around the third axis by the overall angle $2\pi
$ as we move from $-\infty $ to + $\infty $ in the perpendicular direction
(Fig.~\ref{bof}). This can be shown by explicitly evaluating the integral (
\ref{hopf}) for the adiabatic rotation and then by using the fact that the
Hopf term is topological and depends only on the homotopy class of the field
configuration \cite{Wilczek:1983cy}.

Furthermore, it is crucial that $a_\mu$ in Eq.~(\ref{hopf}) coincides up to
a normalization factor with $A_\mu$ introduced in Eq.~(\ref{tue-two}) which
in turn exactly coincides with $\mathcal{A}_\mu^{(3)} = (i/2)\, \mathrm{Tr}
\left( \tau_3\, U^\dagger\,\partial_\mu\, U \right)$, see Eq.~(\ref{cala}).
The latter statement follows from a straightforward calculation which we
present in Appendix. As for the former, let us note that the $z,\,\,\bar z$
fields realize the SU(2) target space. Division by U(1) is implemented
through gauging of U(1), with no kinetic term for the $A_\mu$ field (see
e.g. \cite{ewi}). The following identification
\begin{equation}
a_\mu = \frac{1}{2\pi}\, A_\mu
\end{equation}
ensues \cite{Wu:1984kd,Hlousek:1990du}.

Thus, we conclude that the Hopf invariant is
\begin{equation}
\mathcal{H} = \frac{1}{4\pi ^{2}}\int \, d^{3}x\, \epsilon ^{\mu \nu \rho
}A_{\mu}^{(3)}\partial _{\nu }A_{\rho }^{(3)} = \frac{1}{4\pi ^{2}}\int \,
d^{3}x \, K_0\,.  \label{fri-five}
\end{equation}
The same conclusion was achieved in \cite{Jaroszewicz:1984xw} where the
author used the formalism of the Dirac spinors (cf. Eq.~(\ref{fri-three}))
in a slowly-varying background field configurations which are relevant for
sufficiently wide solitons, and the Goldstone--Wilczek diagrammatic
technique \cite{Goldstone:1981kk}.

Comparing (\ref{fri-four}) with (\ref{fri-five}) we see that the
induced fermion number $F$ for a minimal Hopf Skyrmion is 1. In
general,
\begin{equation}
F = F_0 - \mathcal{H}\,.
\end{equation}
In a related situation the shift of the fermion charge by the Hopf invariant
in the background of the knot soliton was demonstrated in \cite{lfan}.
Although the model of Ref.~\cite{lfan} deals with a single Dirac fermion,
from the purely mathematical standpoint core calculations run in parallel.

At the same time, the current corresponding to the U(1) charge $Q$,
\begin{equation}
J^{\mathrm{U(1)}}_{\alpha\dot\alpha}= \bar{\chi}_{\dot\alpha}\,\tau_3\, {\chi
}_{\alpha }
\end{equation}
is anomaly-free. This implies that there is no induced $Q$ charge.

Now let us return to the issue of the stability of the Hopf Skyrmion and the
hadrons it represents. The theory under consideration contains three kinds
of hadrons whose quantum number assignments are summarized in
Table~\ref{tabtwo}.
\begin{table}[tbp]
\begin{center}
\begin{tabular}{|c|c|c|}
\hline
& \textrm{Charge }$Q$ & $F $\rule{0mm}{5mm} \\ \hline
$\psi $ & $1$ & $1$\rule{0mm}{5mm} \\ \hline
$\pi $ & $2$ & $0$\rule{0mm}{5mm} \\ \hline
\textrm{Skyrmion/exotics} & {$
\begin{array}{c}
0 \\
1
\end{array}
$} & {$
\begin{array}{c}
1 \\
0
\end{array}
$} \rule{0mm}{8mm} \\ \hline
\end{tabular}
\end{center}
\caption{{\protect\footnotesize $Q$ and $F$ mod 2 for nonexotic and exotic
hadrons.}}
\label{tabtwo}
\end{table}
The Hopf Skyrmion can have charges $Q$ and $F$ respectively $0$ and $1$, if
there is no fermion zero mode crossing in the process of evolution from the
topologically trivial background to that of the Hopf Skyrmion, or $1$ and $0$
if there is a fermion zero mode crossing.\footnote{
A relevant discussion of the fermion zero mode crossings in 2+1 dimensions
can be found in \cite{Carena:1990vy}.}

In both cases the lightest exotic hadrons represented by the Hopf Skyrmion
are stable. They can not decay in any number of pions and/or pions plus
``ordinary" baryons with mass $O(N^0)$. Note that this is a $Z_{2}$
stability. Two Hopf Skyrmions can annihilate and decay into an array of $\pi$
's and $\psi$'s. For nonexotic hadron excitations which can be seen in a
constituent model and have mass $O(N^0)$ the combination $Q+F$ is always
even while for exotic hadrons with mass $O(N^2)$ the sum $Q+F$ is odd.

\section{Conclusions}

\label{con}

We considered the low-energy limit of QCD with two adjoint quarks,
summarized by a chiral Lagrangian corresponding to $\chi$SB of the type SU(2)
$\to$U(1). This Lagrangian supports topologically stable solitons described
previously by Faddeev and Niemi. The stability of these solitons is due to
the existence of the Hopf invariant.

We addressed the question of whether one can understand stability of
the corresponding hadrons within the framework of the underlying
microscopic theory. There is a singly strictly conserved quantum
number $Q$ in this theory. We observe, however, that although the
fermion number is not conserved, $(-1)^F$ is. We argued that the
hadrons represented by the Hopf Skyrmions are exotic in the sense
that they have an induced fermion number coinciding with the Hopf
number, so that unlike all ``ordinary" hadrons they are
characterized by negative $(-1)^{Q+F}$. This is the underlying
reason explaining their stability.

Unlike the ``pion" part of the Lagrangian (\ref{lagrangiana})
which is unambiguously fixed by symmetries,
the ``baryon" part (\ref{lagrangianslowlyp})
does not seem to be unique. Indeed, the theory
under consideration has infinitely many fermionic interpolating operators,
besides the one presented in (\ref{fundamentalfermion}).
For example, any operator of the type  Tr$(\lambda \dots\lambda F \dots)$
with an odd
number of $\lambda$'s represents a baryon.
One could pose the question of completeness. We would like to
claim that the Lagrangian
(\ref{lagrangiana}) and  (\ref{lagrangianslowlyp})
is complete in the sense
that it captures all relevant dynamics --- to answer the question we pose
there is no need to inlcude in it additional baryon operators.
The point is that any other baryons with mass $O(N^0)$
(which are necessary unstable particles, resonances)
have a projection on the operator (\ref{fundamentalfermion}).
The inclusion of additional baryon terms would have no impact on our result
since they would not change the anomaly (\ref{anom40}).

There are two obvious questions calling for future investigations:

(i) if one introduces a mass term for the adjoint quarks, how do the Hopf
Skyrmions evolve as the mass term increases?

(ii) when we quantize the Hopf Skyrmions and consider excitations over the
lowest-lying soliton, what is the quantum number assignments for the tower
of the excitations?

\section*{Acknowledgments}

We are grateful to Roberto Auzzi for useful discussions.

This work was supported in part by DOE grant DE-FG02-94ER408. M.S. would
like to thank the Niels Bohr Institute in Copenhagen, where this work began,
for kind hospitality extended to him in July 2006. S.B.~was funded by the
Marie Curie grant MEXT-CT-2004-013510 and in part by FTPI, University of
Minnesota. S.B. wants to thank FTPI for the hospitality in the fall of 2006,
when this work was completed.

\section*{Appendix}

\renewcommand{\theequation}{A.\arabic{equation}} \setcounter{equation}{0}

Let $U(x)$ be a matrix which rotates the  vector  $\vec{n}(x)$ to
align it along the third direction at every given point $x$,
\beq
U^\dagger \, (\vec n\cdot
\vec\tau )\, U =\tau_3\,.
\eeq
Assume that we choose some point $x_0$ and
the corresponding matrix $U(x_0)$. Consider $x_0$ as given and fixed.

In the vicinity of $x_0$,
$$x=x_0 +\delta x,$$
the matrix $U(x)$ can be decomposed in two factors
\begin{equation}
U(x)=U(x_0)\, \delta U(\delta x) \, ,
\end{equation}
where the space-time dependence is now enclosed only in $\delta
U(\delta x)$. Since $\delta U(0)={\bf 1}$, we can perform  the expansion
$\delta U(\delta x)=\exp\left\{ i\left(\nu_1\tau_1
+\nu_2\tau_2\right) \right\}$ where $\nu_{1,2}$ depend on $\delta
x$. We will expand the exponent keeping the leading relevant terms.
Then
\begin{eqnarray}
n_1 &=& -2 \nu_2\,,\quad n_2 = 2\nu_1\,,  \notag \\[2mm]
n_3 &=& 1-2\left(\nu_1^2+\nu_2^2\right)\,.
\end{eqnarray}
The two-component $z$ appearing in the gauged definition of the sigma model
is
\begin{equation}
z = \left(
\begin{array}{c}
\varepsilon \\[2mm]
1-\frac{1}{2}|\varepsilon |^2
\end{array}
\right)\,,\qquad \varepsilon =\nu_2 +i\nu_1\,.
\end{equation}
The vector $A_\mu$ defined in Eq.~(\ref{tue-two}) takes the form
\begin{equation}
A_\mu = - \nu_1 \,\overset{\leftrightarrow}{\partial}_\mu\, \nu_2\,.
\end{equation}
At the same time, the vector $\mathcal{A}$ defined in Eq.~(\ref{cala}) is
\begin{equation}
\mathcal{A}_\mu = i\,U^\dagger\partial_\mu\,U = - \tau_3 \left( \nu_1 \,
\overset{\leftrightarrow}{\partial}_\mu\, \nu_2\right) +\mathrm{terms \,\,\,
with\,\,\,} \tau_{1,2}\,.
\end{equation}
Thus we have $A_\mu = \mathcal{A}_\mu^{(3)}$ that, in the original
frame, means
\begin{equation} A_\mu = \mathcal{A}_\mu^{(i)}n^{(i)}\,.
\end{equation}

\end{document}